\documentclass[submission, Proceedings]{SciPost}

\hypersetup{
    colorlinks,
    linkcolor={red!50!black},
    citecolor={blue!50!black},
    urlcolor={blue!80!black}
}

\usepackage[bitstream-charter]{mathdesign}
\usepackage{hyperref}
\DeclareSymbolFont{usualmathcal}{OMS}{cmsy}{m}{n}
\DeclareSymbolFontAlphabet{\mathcal}{usualmathcal}

\begin{document}

\begin{center}{\Large \textbf{
Sub-GeV Dark Matter Searches with EDELWEISS: New results and prospects
\\
}}\end{center}

\begin{center}
H. Lattaud\textsuperscript{1*} for the EDELWEISS collaboration
\end{center}

\begin{center}
{\bf 1} Univ Lyon, Universit\'e Lyon 1, CNRS/IN2P3, IP2I-Lyon, F-69622, Villeurbanne, France
* lattaud@ip2i.in2p3.fr
\end{center}

\begin{center}
\today
\end{center}


\definecolor{palegray}{gray}{0.95}
\begin{center}
\colorbox{palegray}{
  \begin{tabular}{rr}
  \begin{minipage}{0.1\textwidth}
    \includegraphics[width=30mm]{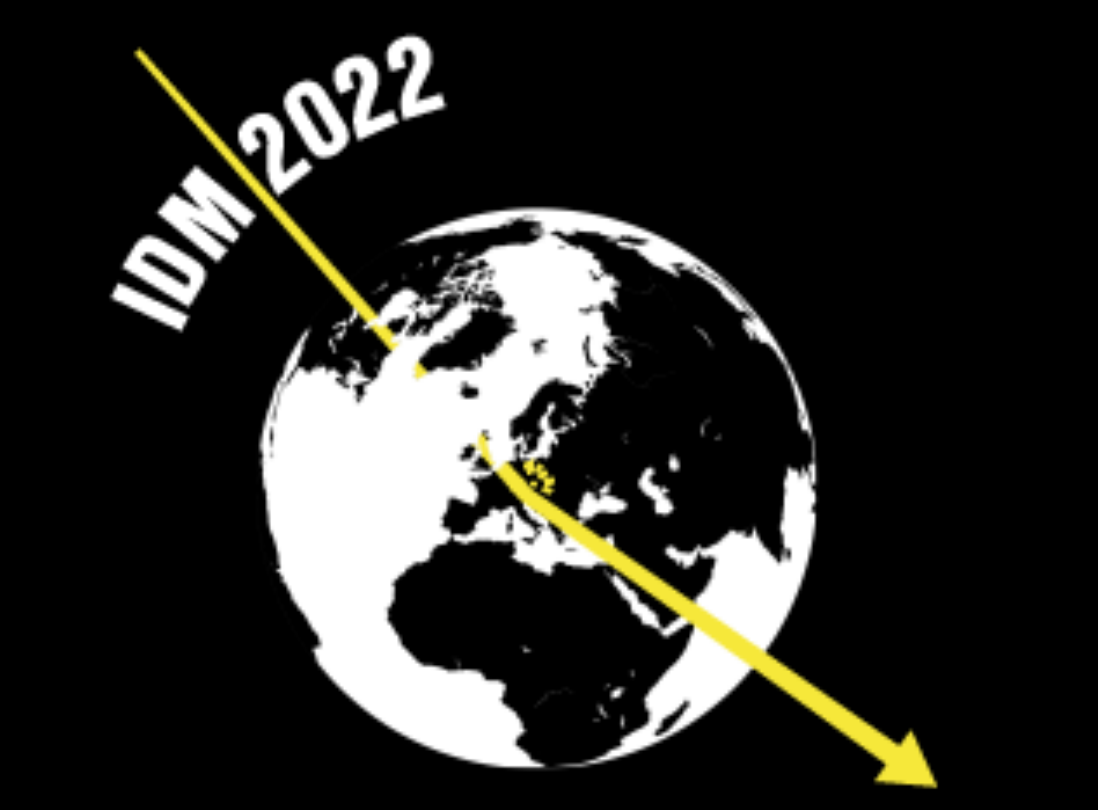}
  \end{minipage}
  &
  \begin{minipage}{0.85\textwidth}
    \begin{center}
    {\it 14th International Conference on Identification of Dark Matter}\\
    {\it Vienna, Austria, 18-22 July 2022} \\
    https://doi.org/{10.21468/SciPostPhysProc.?}\\
    \end{center}
  \end{minipage}
\end{tabular}
}
\end{center}

\section*{Abstract}
{\bf
The Edelweiss collaboration performs light Dark Matter (DM) particles searches with germanium bolometer collecting
charge and phonon signals. Thanks to the Neganov-Trofimov-Luke (NTL) effect, a RMS resolution of 4.46 electron-hole pairs
was obtained on a massive (200g) germanium detector instrumented with a NbSi Transition Edge Sensor (TES) operated underground at the Laboratoire Souterrain de Modane (LSM).
This sensitivity made possible a search for WIMP using the Migdal effect down to 32 MeV/C$^{2}$ and exclude cross-sections down to 10$^{-29}$ cm$^2$.
It is the first measurement in cryogenic germanium with such thermal sensor, proving the high relevance of this technology.
Furthermore, such TES have shown sensitivity to out-of-equilibrium phonons, paving the way for EDELWEISS new experience CRYOSEL.
This is an important step in the development of Ge detectors with improved performance in the context of the EDELWEISS-SubGeV program.
}

\vspace{10pt}
\noindent\rule{\textwidth}{1pt}
\tableofcontents\thispagestyle{fancy}
\noindent\rule{\textwidth}{1pt}
\vspace{10pt}

\section{Introduction}
\label{sec:intro}

There were great progresses in the Dark Matter (DM) direct interaction with nuclei searches \cite{Drukier:1984vhf,Goodman:1984dc,Drukier:1986tm}, for particle with masses ranging from 1 GeV$\cdot$c$^{-2}$ to $1$~TeV$\cdot$c$^{-2}$~\cite{xenon1t,lux,pandax}. Nevertheless, because of the absence of signal in that region, the interest in extending the search to lower masses intensified ~\cite{Essig,Cheung,Hooper,Falkowski,Petraki,Zurek,Bertone:2018krk}. Probing lower masses raise additional experimental constraints, such as the need of lower energy threshold (below 1~keV) as well as ionization or scintillation yield for nuclear recoil. To such constraints, collaborations have to add dealing with new types of low energy backgrounds ~\cite{cresst,cdmslite,red30,damic,sensei,cedex}. 

In order to avoid the issues raised by the very low kinetic energy of the nuclear recoil, it is possible to use the Migdal effect  ~\cite{Vergados:2004bm,Moustakidis:2005gx,Bernabei:2007jz,Ibe:2017yqa}.
This effect quantifies the probability that an atomic electron is released because of the DM particle scattering on a nucleus. The  electrons are emitted with an energy typically much larger than the kinematic energy of the nuclear recoil ~\cite{Ibe:2017yqa}, yielding signals much easier to detect. 

In  this context, the EDELWEISS collaboration used the Migdal effect to extend the mass range of DM particles search   down to $32$~MeV$\cdot$c$^{-2}$, using a $200$~g cryogenic Ge detector equipped  with a NbSi Transition Edge Sensor (TES) and operated underground at the Laboratoire Souterrain de Modane (LSM). This detector achieved a low energy threshold for electron recoil and appeared to be less sensitive to the Heat Only background (HO), an already known background  made of events that are not associated with charges. 

The development of the NbSi detector is part of the EDELWEISS-SubGeV program and a step toward the new generation of EDELWEISS detectors, Cryosel. Those detectors will be  Ge bolometers able to sustain high biases and equipped with a new sensor able to tag down to a single charge and thus veto HO events \cite{Marnieros:2022bsi}.

\section{New results : DM search with a detector equipped with TES  } \label{sec:results}

The search was performed at the LSM benefitting from the ultra-low background environment of the EDELWEISS-III cryostat~\cite{edwtech} and the 4800 m.w.e. rock overburden. The experiment, its results and their interpretation are described in detail in \cite{NbSi209Migdal}. The detector named NbSi209 is a $200$~g Ge cylindrical crystal (48~mm in diameter and $20$~mm in height) on top of which was lithographed a Nb$_x$Si$_{1-x}$ thin film TES~\cite{nbsi-ltd} used as thermal sensor. The TES is operated around  44 mK, the temperature at which the transition between the normal and the superconducting states start to occur. 
The data from the ionization and heat channels were digitized at a frequency of $100$~kHz, filtered, and continuously stored on disk with a digitization rate of $500$~Hz. 
The bottom side is fully covered by an electrode biased at a voltage varying between $\pm 66$~V.
The bias is applied in order to trigger the so-called Neganov-Trofimov-Luke (NTL) effect\cite{Neganov,Luke}.
This effect states that the drift of the  $N$ electron-hole pairs across a voltage difference $\Delta V $ creates additional phonons of energy $E_{NTL}$ =  $Ne\Delta V$ ($e$ is the elementary charge). This energy 
$E_{NTL}$  adds to the recoil energy. In the case of events associated with charges, the mean gain of the NTL effect is  $\langle g \rangle =1+e\Delta V /\epsilon_{gamma}$, where $\epsilon_{gamma} = 3.0$~eV is the average ionization energy in Ge for electron recoils~\cite{knoll}. The average resolution over the dataset is  between $90$ and $100$~eV. For the NTL gain of $\langle g \rangle=23$ corresponding to a bias of $66$~V, this leads to a resolution of approximately $4$~eV electron-equivalent (eV$_{ee})$.

\begin{figure}[!h]
\includegraphics[width=0.5\linewidth]{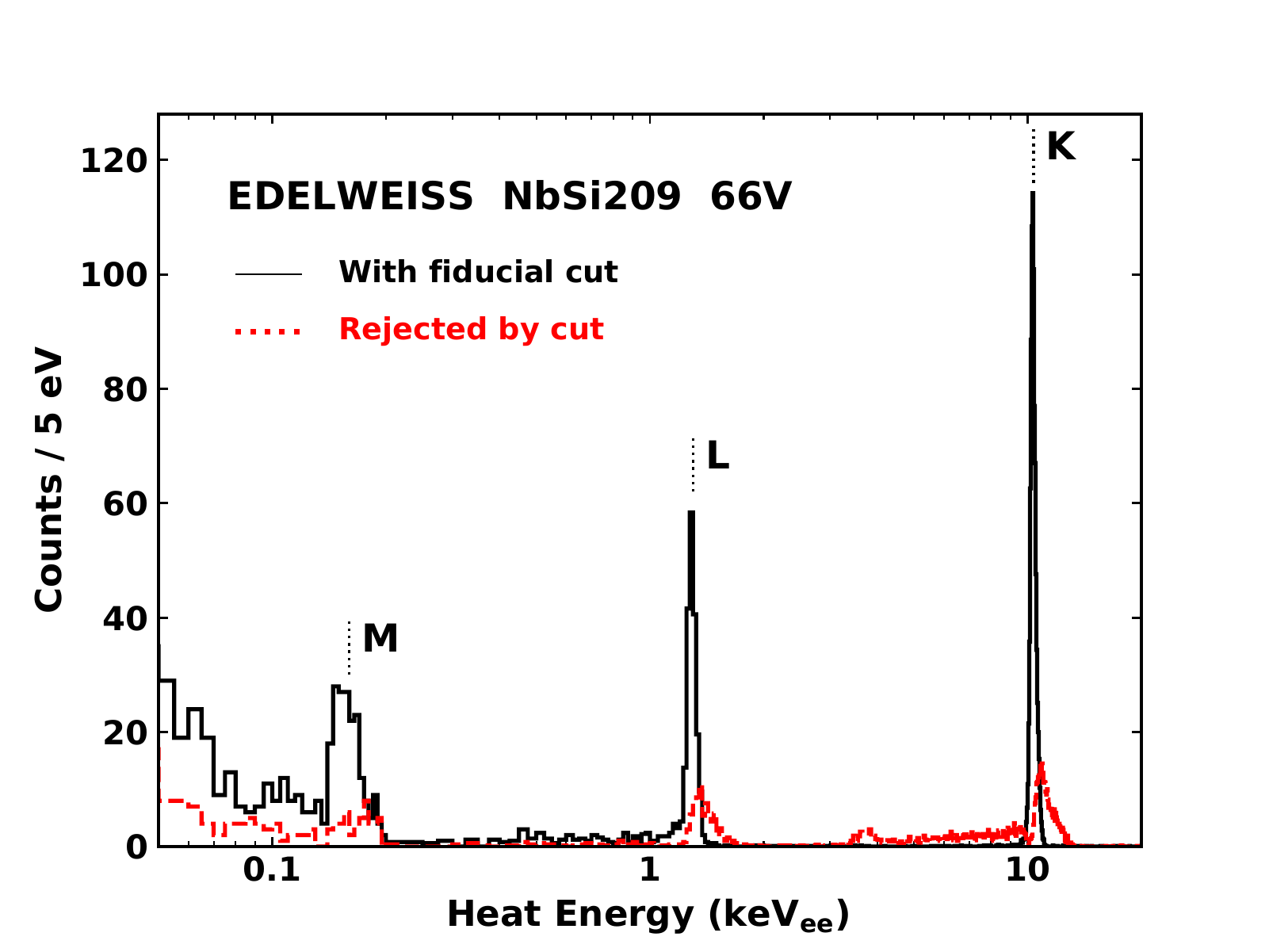}
\includegraphics[width=0.5\linewidth]{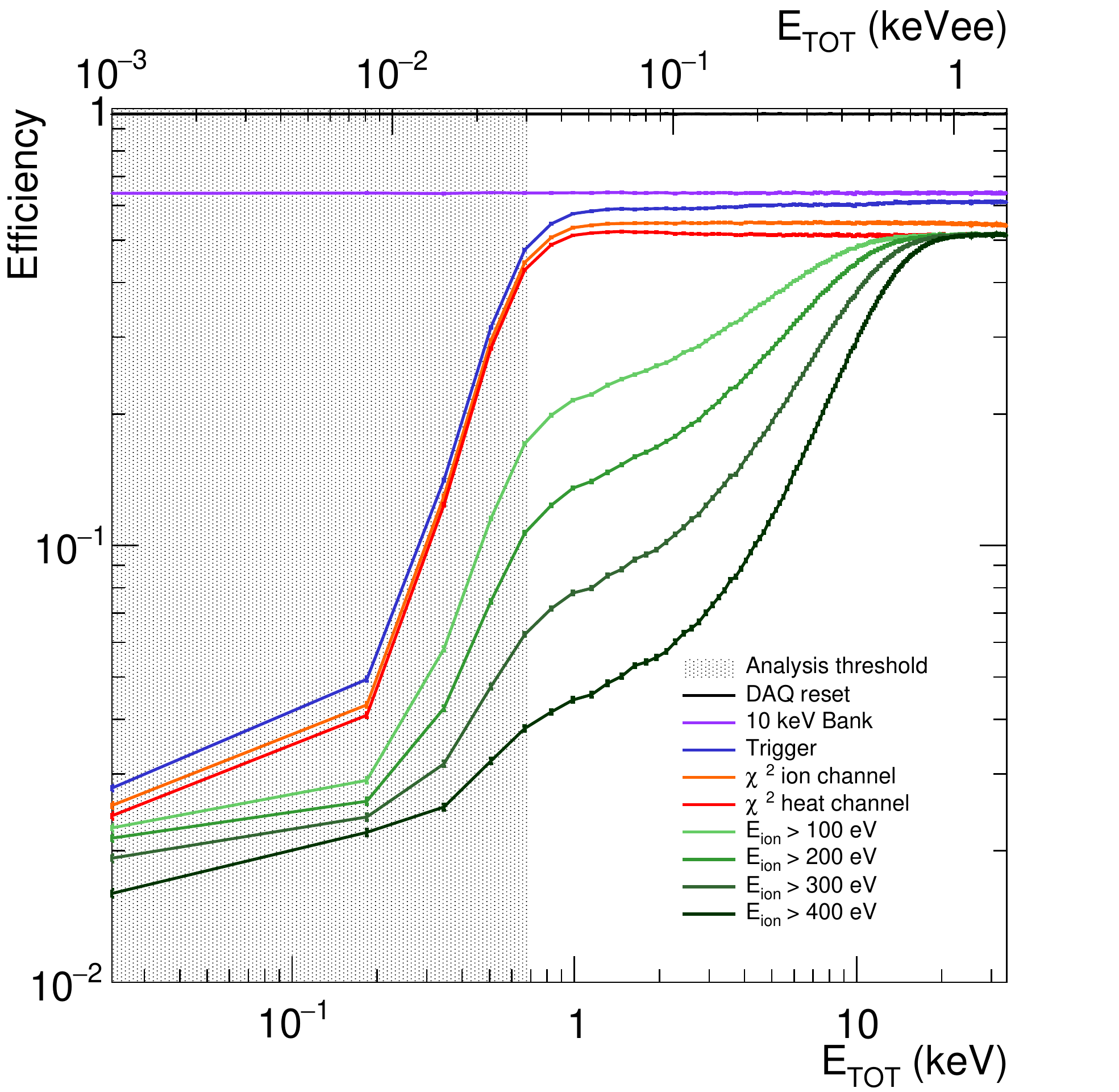}

\caption{Left: energy spectrum recorded at a bias of $66$~V following  the $^{71}$Ge emitted x-ray lines induced by the neutron activation of the Ge detector, resulting in the strong lines characteristic of $^{71}$Ge x-ray emission. Right :  efficiency as a function of the input phonon heat energy of simulated events inserted in the data stream in units of keV (lower axis), or keV-electron-equivalent (keV$_{ee}$, upper axis). Each color line corresponds to a step in the trigger and data analysis process. Those figures have been taken from \cite{NbSi209Migdal} under RNP/22/OCT/058564 license. }\label{fig:nbsi_picture}
\end{figure}

The detector was activated with a strong AmBe neutron  source. This produces  short-lived $^{71}$Ge isotopes which decay by electron capture in the K, L, and M shells, with de-excitation x-ray lines at $10.37$, $1.30$, and $0.16$~keV, respectively. The isotopes being produce uniformly throughout the detector volume, the absorption  of the low energy x-ray lines allows to probe the DM response in the entire detector volume, providing a precise characterization of low energy calibration and data selection efficiency. Those lines are visible in Fig.~\ref{fig:nbsi_picture}, left panel, which shows the energy spectrum for the phonon signal from calibration data recorded at a bias of $66$~V. 
A sample of $10667$ K-shell events with  energies between $2$ and $12.6$~keV$_{ee}$, recorded at $66$~V was used to assess the  efficiency of data processing, triggering and selection as function of the energy. In order to probe the energy
dependency, those events were scaled to the desired energy and injected at random time in the data
stream. Those events went through the same analysis pipeline as the original data. The Figure~\ref{fig:nbsi_picture}, right panel shows the efficiency as a function of the energy of the injected pulse at various steps of the triggering, processing and selection of the data analysis. This procedure of event simulation accounts for actual noise conditions in data, as well as reconstruction and selection biases. The efficiency plateau is around $\sim 50\%$ for the bulk of the selection, in top of which a stringent criterion on the ionization energy is added in order to reject most of the HO background, the effect on efficiency is shown by the green curves.

\begin{figure}[!h]
\includegraphics[width=0.5\linewidth]{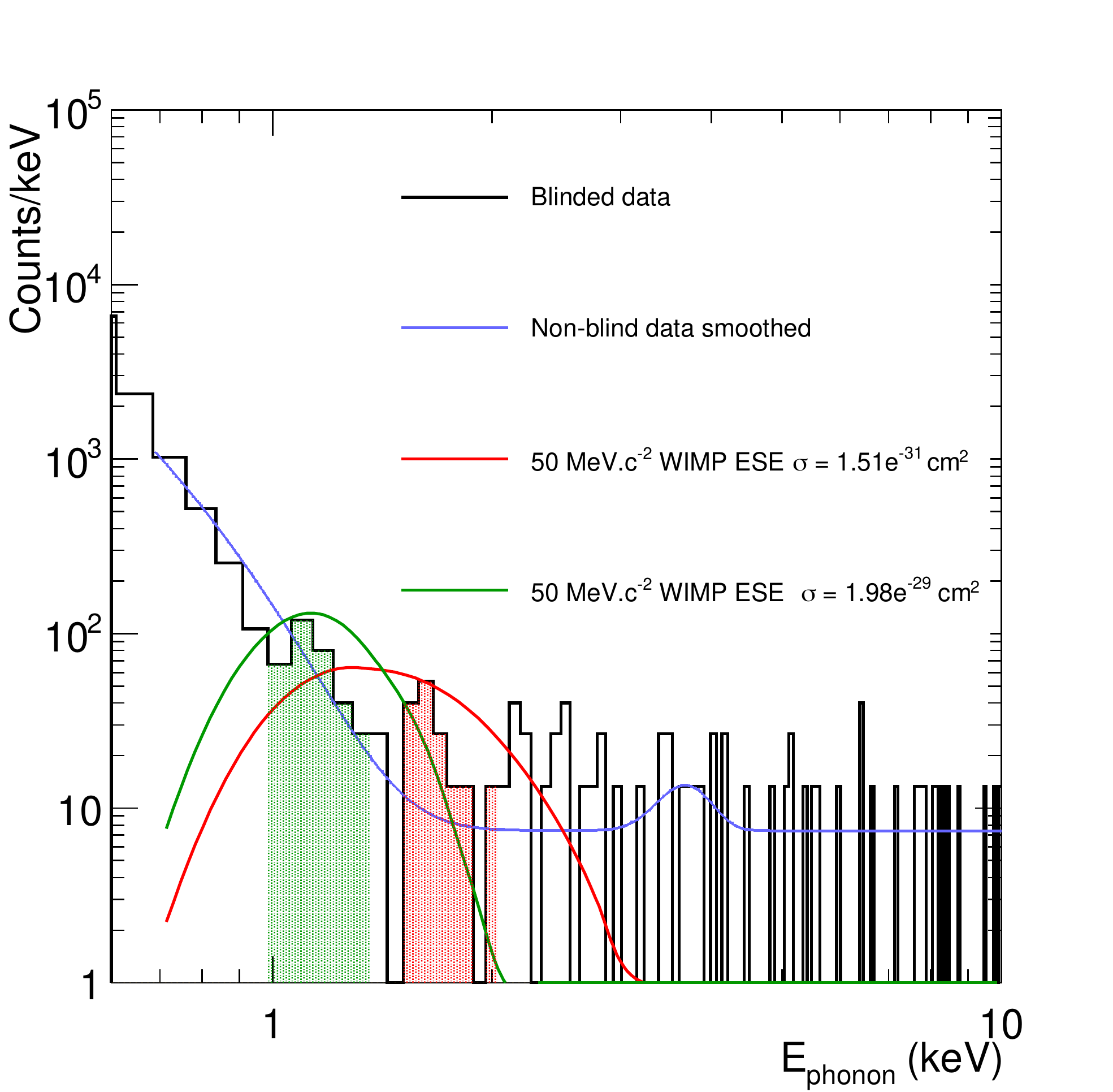}
\includegraphics[width=0.55\linewidth]{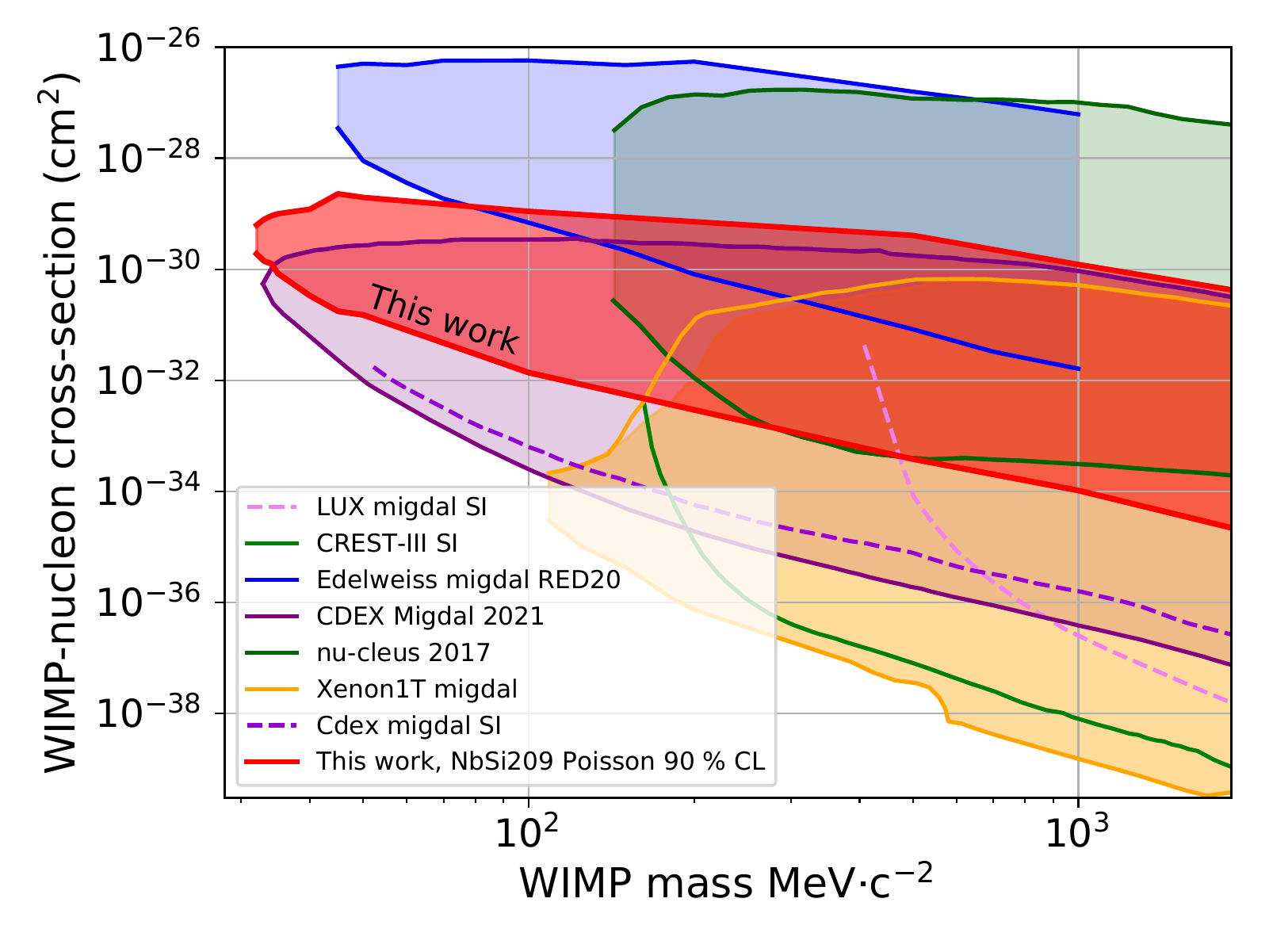}

\caption{Left: energy spectrum after all criteria are applied for the blinded dataset in counts per keV (black histogram). Reference sample data smoothed by analytical function (plain blue). The red and green curves show the  excluded  Migdal spectra smeared to detector resolution, corrected for the Earth-shielding effect and efficiency corrected  for  WIMPs of $50$~MeV$\cdot$c$^{-2}$ for lower and upper excluded cross-section. Right : $90\%$~C.L. upper limit on the cross-section for Spin-Independent interaction  between DM and Ge nuclei through the Migdal effect. The red contour is obtained by taking into account the slowing of the DM particle flux through the material above the detector. These results are compared to other experiments \cite{Akerib:2018hck,nucleus,cresst3,xenon-migdal,cedex,cedex2021}. Those figures have been taken from \cite{NbSi209Migdal} under RNP/22/OCT/058564 license.}\label{fig:results}
\end{figure}

The data have been interpreted in terms of  limits on the spin-independent interaction of DM particles with target atoms through the so-called Migdal effect using calculations from \cite{Ibe:2017yqa,outer-essig}. In addition, because of the underground location of the detector, the action of the stopping power of the rock overburden on the DM particle flux has been taken into account \cite{Kavanagh:2016pyr, Emken:2017qmp, Mahdawi:2018euy, Hooper:2018bfw}. The search was performed with data recorded between March 2019 and June 2020, only considering samples with a baseline heat energy resolution lower than $140$~eV RMS. The resulting average resolution is 102 $\pm$ 12~eV RMS, corresponding to  4.46 $\pm$ 0.54~eV$_{ee}$ RMS once the NTL gain $\langle g \rangle$ is considered. Half of the dataset was blinded to perform the search, the other half was used to set the analysis criteria. The number of events in the chosen energy range (the regions of interest ROI's), is used to set the $90 \%$~C.L.  Poisson upper limit on DM particle interacting through Migdal effect with Ge nuclei. The signal energy distributions for a WIMP of $50$~MeV, corrected for efficiency and smeared to detector resolution for upper and lower excluded cross-sections are shown in Fig~\ref{fig:results} left panel in green and red, respectively, as well as their corresponding ROI's. The non-blinded sample was used to derive a data model in order to set the ROI's prior to unblind the rest of the data. This model corresponds to the blue curve in Fig~\ref{fig:results} left panel.

The resulting limits are shown by the red contour (delimited by the thick red line) in Fig~\ref{fig:results}, right panel. Those DM constraints go down to  $32$~MeV$\cdot$c$^{-2}$. Below that mass, the large cross-section that would yield an observable signal is prevented by the  stopping effect of the overburden and shielding. This contour is compared with results from other experiments \cite{Akerib:2018hck,nucleus,cresst3,xenon-migdal,cedex,cedex2021} and with the previous EDELWEISS Migdal search  performed above ground. This shows orders of magnitude of improvement with this previous search  and that the underground operation of the detector did not jeopardize the potential of a Migdal search.

\begin{figure}[!h]
\begin{center}
    
\includegraphics[width=0.5\linewidth]{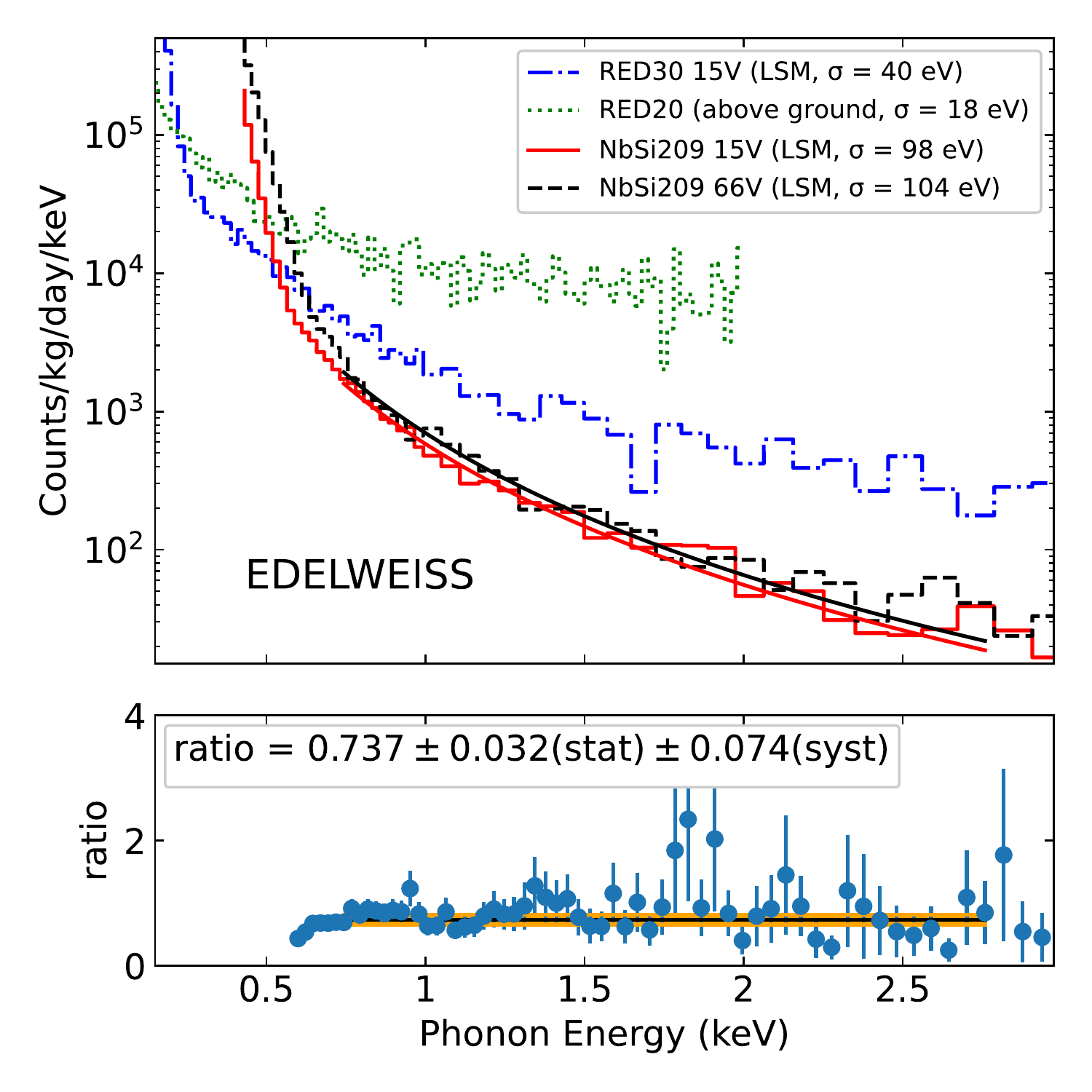}
\caption{Heat energy spectra of events recorded with RED20 operated above ground at $0$~V (green)\cite{red20}, events with  no ionization ($E_{ion} < 0$) for the RED30 detector operated at $15$~V (blue),  NbSi209 operated at $15$~V (red) and $66$~V (black). The fitted power law on NbSi spectra when operated at  $66$~V and $15$~V in black and red respectively. The lower figure shows the ratio of NbSi209 distributions recorded at $15$ and $66$~V (blue) and the associated fit of a constant (black line) and its uncertainty band  (orange). This figure has been taken from \cite{NbSi209Migdal} under RNP/22/OCT/058564 license.}\label{fig:chargelimits}
\end{center}

\end{figure}

Nevertheless, despite the efforts made to reject the HO contamination, this analysis was background limited. Although its HO nature is not questionable at high energy because of the ionization signal associated with the heat signal, the ionization resolution of $\sim 200$~eV (RMS) prevent such conclusion at lower energy.
Fig~\ref{fig:chargelimits} shows energy distribution of data recorded at 15~V and 66~V in red and black, respectively. The lack of quantitative shape differences of those spectrums is a hint of their HO nature. The fit of a power law ($\alpha E ^ {\beta}$) yields identical slopes within uncertainties $\beta \sim 3.40$ for both spectra.
The compatible fitted powers indicate that no NTL amplification occurs, and therefore, that no charges are involved. The flatness of the ratio shown on the bottom pad enhanced this observation. Furthermore, assuming the that both the HO background and the possible electronic background follow the same power law, a limit on the fraction of events associated with charges $x$ can be set, the resulting upper limit is $x<0.0004$ at 90\%C.L. This confirms that the nature of this background is mainly HO in the energy range of the search between $0.8$ to $2.8$~keV.

\section{Prospect : Cryosel  } \label{sec:prospect}
\begin{figure}[!h]
\begin{center}
\includegraphics[width=0.5\linewidth]{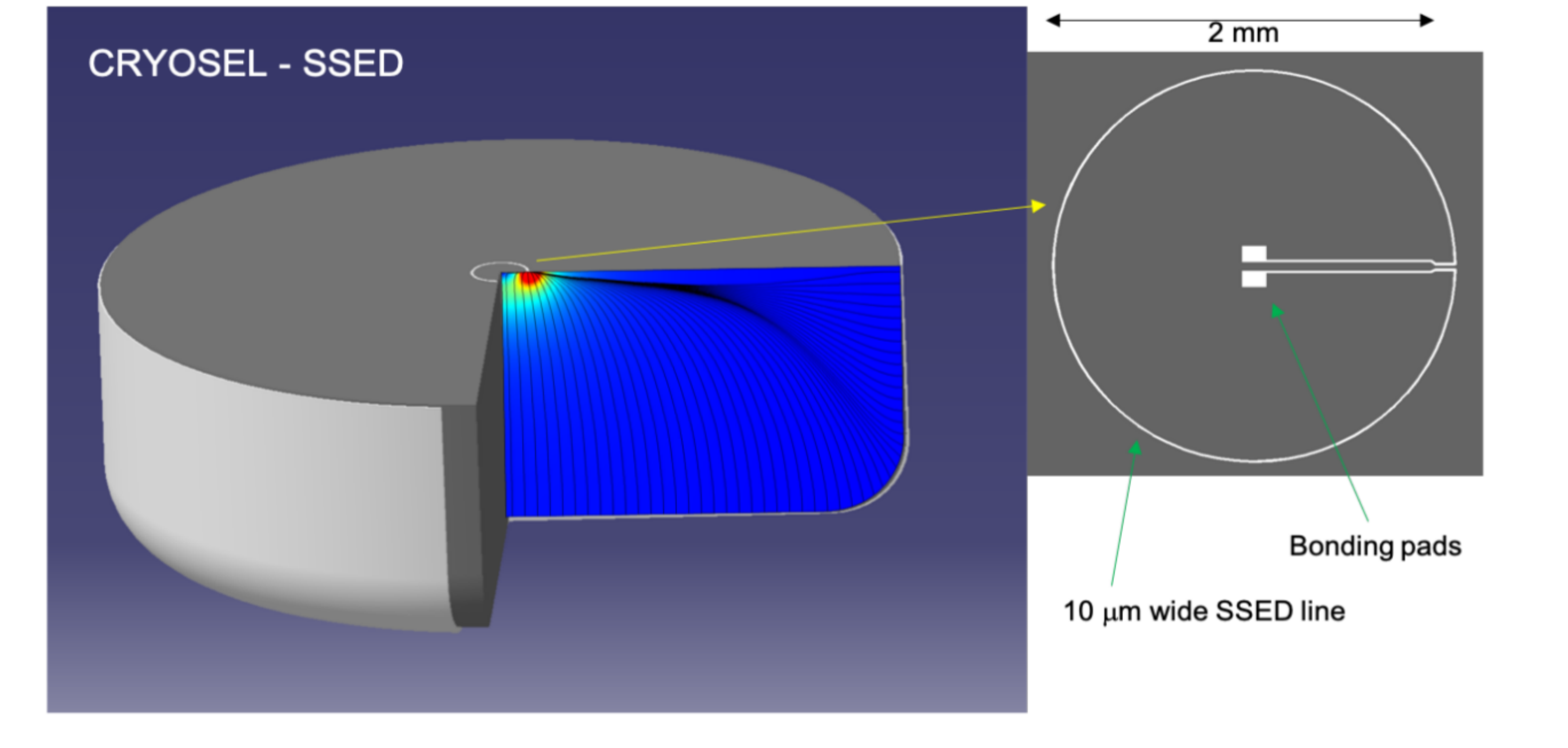}    
\includegraphics[width=0.45\linewidth]{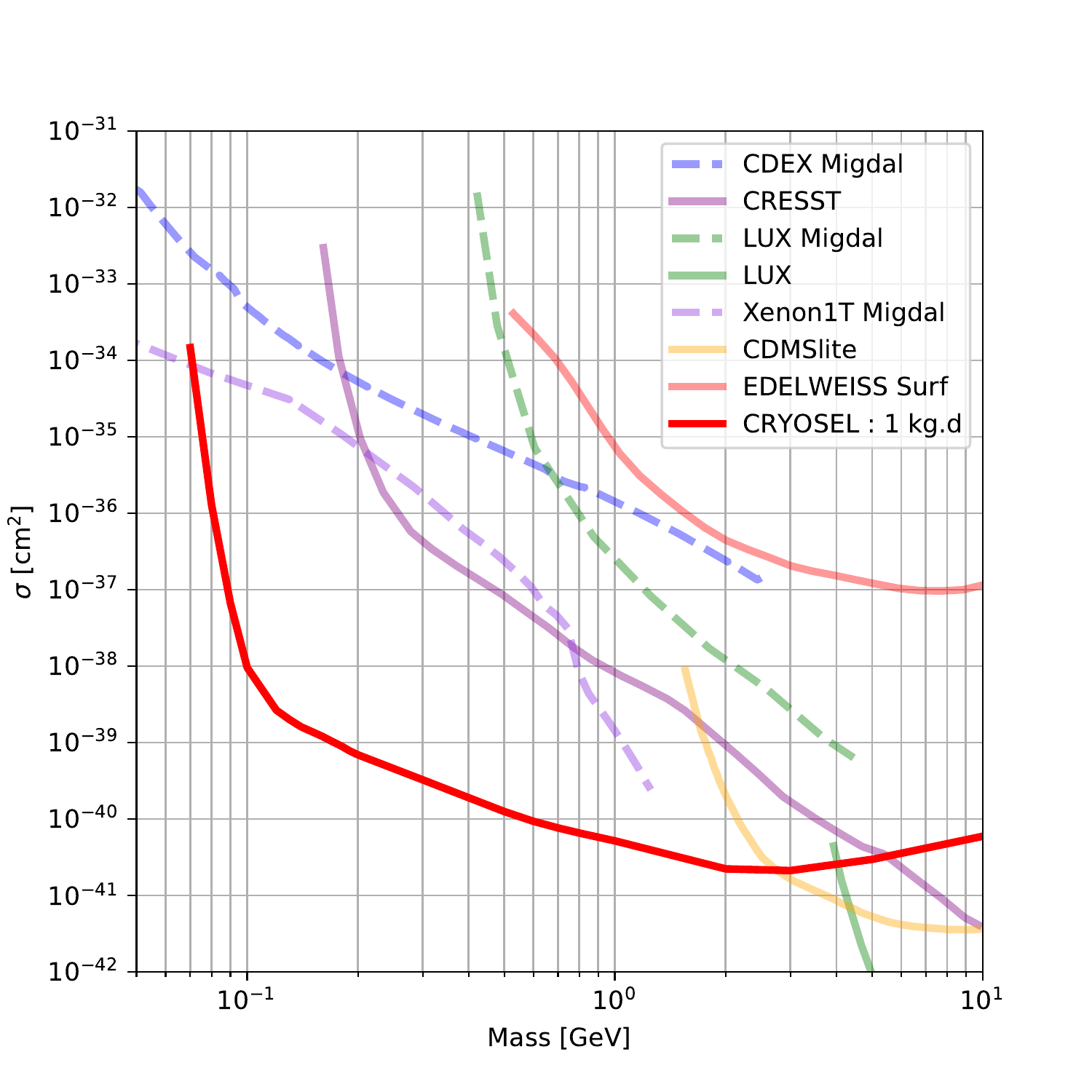}
\caption{Left: Cryosel 30 g Ge cryogenic detector with covering bottom Al electrode and the SSED detector layout including the simulated electric field-lines. The intensity of the electric field rises significantly
under the SSED line (red region)
Right: thick red line: projected sensitivity for WIMP-nucleon spin-independent interaction ($90\%$CL) with a 1 kg.d exposure of the CRYOSEL detector ($\sim 1$ month with a 30 g detector). These results are compared to other experiments \cite{red20,cdmslite,cedex,cresst3,xenon-migdal,Akerib:2018hck,lux}}\label{fig:cryosel}
\end{center}
\end{figure}
The next step in the EDELWEISS Sub-GeV program is to improve both the phonon resolution
of the detectors and the maximum bias at which they can be operated. Developments are ongoing
on the study and reduction of the HO background, as it is one of the limiting factor in future
searches in EDELWEISS.  For example, a new detector with a novel phonon sensor, consisting of a 10 $\mu$m thin  NbSi line called Super Conducting Single Electron Device (SSED)  evaporated  on top of a 40 g Ge crystal is under study. Such sensor  on  a detector operated at high biases ($\sim 200$~V) has a sensitivity to the athermal phonons triggered by the NTL effect of a single charge drifting in the strong electric field.
This leads to the possibility of evaluating the purely thermal nature of HO events and thus vetoing this background. The goal of CRYOSEL is to successfully couple this SSED with a high performance bolometer with 20 eV (RMS) of baseline resolution. The high-resolution thermal measurement is obtained by using a NTD-Ge thermistor with 20~eV phonon resolution and  applying a bias of 200~V on the detector. This design is shown in the left panel of Fig~\ref{fig:cryosel} along with the structure of the expected electric field.
The right panel shows the expected sensitivity of such  design to WIMP nuclei interaction, this would be competitive with the performances of much more massive detectors and cover areas of the parameters space that have yet to be excluded. It would also cover all the milestone models \cite{milestonemodel} for DM interacting with electron. This program is in it R\&D stage and prototypes are currently being produced and tested.

\section{Conclusion}
The EDELWEISS collaboration has searched for DM particle interaction exploiting the Migdal effect with WIMP masses between $32$~MeV$\cdot$c$^{-2}$ and $2$~GeV$\cdot$c$^{-2}$ using a $200$~g Ge detector operated underground at the LSM. The search constrains a new region of the parameters space 
for cross-sections close to $10^{-29}$~cm$^2$. In the context of its EDELWEISS-SubGeV program, the collaboration is also investigating new methods to significantly reduce HO backgrounds by improving its ionization resolution with the use of new cold preamplifiers~\cite{hemt}, and by developing NbSi-instrumented devices able to tag the out-of-equilibrium NTL phonons associated to a single electron.

\paragraph{Funding information}
The EDELWEISS project is supported in part by the French Agence Nationale pour la Recherche (ANR) and the LabEx Lyon Institute of Origins (ANR-10-LABX-0066) of the Universit\'e de Lyon within the program ``Investissements d'Avenir'' (ANR-11-IDEX-00007), by the P2IO LabEx (ANR-10-LABX-0038) in the framework ``Investissements d'Avenir'' (ANR-11-IDEX-0003-01) managed by the ANR (France), and the Russian Foundation for Basic Research (grant No. 18-02-00159). This project  has  received  funding  from  the  European Union’s Horizon 2020 research and innovation program under the Marie Sk\l odowska-Curie Grant Agreement No. 838537. B.J. Kavanagh thanks the Spanish Agencia Estatal de Investigaci\'on (AEI, MICIU) for the support to the Unidad de Excelencia Mar\'ia de Maeztu Instituto de F\'isica de Cantabria, ref. MDM-2017-0765.
We thank J.P. Lopez (IP2I) and the Physics Department of Universit\'{e} Lyon 1 for their contribution to the radioactive sources.




\bibliographystyle{SciPost_bibstyle}

\begin{thebibliography}{100}

\bibitem{Goodman:1984dc}
  Goodman, Mark W. and Witten, Edward
  \href{https://doi.org/10.1103/PhysRevD.31.3059}{Phys.\ Rev.\ D.\ {\bf 31}, 3059 (1985) }. 

\bibitem{Drukier:1986tm}
A. Drukier, K. Freese, D. N. Spergel
\href{https://doi.org//10.1103/PhysRevD.33.3495}{Phys.\ Rev.\  D.\ {\bf33}, 3495-3508 (1986).}

\bibitem{Drukier:1984vhf} 
A. Drukier, L. Stodolsky,  
 \href{https://doi.org//10.1103/PhysRevD.30.2295}{Phys.\ Rev.\ D.\ {\bf 30}, 2295 (1984)}. 

\bibitem{xenon1t} 
E. Aprile {\it et al.} (XENON Collaboration),
\href{https://doi.org//10.1103/PhysRevLett.121.111302}{Phys.\ Rev.\ Lett.\ {\bf 121}, 111302 (2018)}, arXiv:1805.12562.

\bibitem{lux} 
D. S. Akerib {\it et al.} (LUX Collaboration),
\href{https://doi.org//10.1103/PhysRevLett.118.021303}{Phys.\ Rev.\ Lett.\ {\bf 118}, 021303 (2017)}, arXiv:1608.07648.


\bibitem{pandax} 
A. Tan {\it et al.} (PandaX-II Collaboration),
\href{https://doi.org//10.1103/PhysRevLett.117.121303}{Phys.\ Rev.\ Lett.\ \textbf{117}, 121303 (2016)}, arXiv:1607.07400.

\bibitem{Essig} 
R. Essig, J. Kaplan, P. Schuster and N. Toro,  (2010),
arXiv:1004.0691.

\bibitem{Cheung} 
C. Cheung, J. T. Ruderman, L.-T. Wang and I. Yavin,
\href{https://doi.org//10.1103/PhysRevD.80.035008}{Phys.\ Rev.\ D {\bf 80}, 035008 (2009)}, arXiv:0902.3246.

\bibitem{Hooper} 
D. Hooper and W. Xue, 
\href{https://doi.org//10.1103/PhysRevLett.110.041302}{Phys.\ Rev.\ Lett.\ {\bf 110}, 041302 (2013)}, arXiv:1210.1220.

\bibitem{Falkowski} 
A. Falkowski, J.T. Ruderman and T. Volansky,
\href{https://doi.org//10.1007/JHEP05(2011)106}{JHEP {\bf 1105}, 106 (2011)}, arXiv:1101.4936.

\bibitem{Petraki} 
K. Petraki and R.R. Volkas, 
\href{https://doi.org//10.1142/S0217751X13300287}{Int. J. Mod. Phys. A {\bf 28}, 1330028 (2013)}, arXiv:1305.4939.

\bibitem{Zurek} 
K. M. Zurek, 
\href{https://doi.org//10.1016/j.physrep.2013.12.001}{Phys.\ Rep.\ {\bf 537}, 91 (2014)}, arXiv:1308.0338.

\bibitem{Bertone:2018krk} 
G.~Bertone and T.~M.~P.~Tait,
  \href{https://doi.org//10.1038/s41586-018-0542-z}{Nature {\bf 562}, 51-56 (2018)}, arXiv:1810.01668.

\bibitem{red30}
Q. Arnaud {\it et al.} (EDELWEISS Collaboration),
\href{10.1103/physrevlett.125.141301}{Phys.\ Rev.\ Lett.\ \textbf{135}, 141301 (2020)},
arXiv:2003.01046


\bibitem{cresst} 
G. Angloher {\it et al.} (CRESST Collaboration), \href{https://doi.org//10.1140/epjc/s10052-016-3877-3}{Eur.\ Phys.\ J.\ C.\ {\bf 76}, 25 (2016)}, arXiv:1509.01515.

\bibitem{cdmslite} 
R. Agnese {\it et al.} (SuperCDMS Collaboration),
\href{https://doi.org//10.1103/PhysRevD.97.022002}{Phys.\ Rev.\ D {\bf 97}, 022002 (2018)}, arXiv:1707.01632.

\bibitem{damic}
A. Aguilar-Arevalo {\it et al.} (DAMIC Collaboration),
\href{10.1103/PhysRevLett.118.141803}{Phys.\ Rev.\ Lett.\ {\bf 118}, 141803 (2017)}, arXiv:1611.03066.

\bibitem{sensei}
O. Abramoff {\it et al.} (SENSEI Collaboration),
\href{10.1103/physrevlett.122.161801}{Phys.\ Rev.\ Lett.\ {\bf 122}, 161801 (2019)}, arXiv:2004.11378.

\bibitem{cedex}
Z.Z. Liu {\it et al.} (CDEX Collaboration),
\href{10.1103/physrevlett.123.161301}{Phys.\ Rev.\ Lett.\ {\bf 123}, 161301 (2019)} arXiv:1905.00354.



\bibitem{Vergados:2004bm} 
J.~D.~Vergados and H.~Ejiri,
\href{https://doi.org//10.1016/j.physletb.2004.11.085}{Phys.\ Lett.\ B {\bf 606}, 313 (2005)}, arXiv:hep-ph/0401151.



\bibitem{Moustakidis:2005gx} 
C.~C.~Moustakidis, J.~D.~Vergados and H.~Ejiri,
\href{https://doi.org//10.1016/j.nuclphysb.2005.08.033}{Nucl.\ Phys.\ B {\bf 727}, 406 (2005)}, arXiv:hep-ph/0507123.

\bibitem{nucleus} 
G.~Angloher {\it et al.} (CRESST Collaboration),
\href{https://doi.org//10.1140/epjc/s10052-017-5223-9}{Eur.\ Phys.\ J.\ C {\bf 77}, 637 (2017)}, arXiv:1707.06749.

\bibitem{SuperCDMS:2020aus}
D.W.~Amaral~{\it~et~al.}
\href{https://doi.org//10.1103/PhysRevD.102.091101}{Phys.\ Rev.\ D.\ {\bf 102}, 091101(2020)}

\bibitem{red20}
E. Armengaud {\it et al.} (EDELWEISS Collaboration),
\href{10.1103/physrevd.99.082003}{Phys.\ Rev.\ D \textbf{99}, 082003 (2019)},
arXiv:1901.03588

\bibitem{cedex2021}
Z.Z. Liu {\it et al.} (CDEX Collaboration),
arXiv:2111.11243.


\bibitem{Bernabei:2007jz} 
R.~Bernabei {\it et al.},
  \href{https://doi.org//10.1142/S0217751X07037093}{Int.\ J.\ Mod.\ Phys.\ A {\bf 22}, 3155 (2007)}, arXiv:0706.1421.

\bibitem{Ibe:2017yqa} 
M.~Ibe, W.~Nakano, Y.~Shoji and K.~Suzuki,
  \href{https://doi.org//10.1007/JHEP03(2018)194}{JHEP {\bf 1803}, 194 (2018)}, arXiv:1707.07258.

\bibitem{Marnieros:2022bsi}
Marnieros, S. and others, 2022
    arXiv:2201.01639
    
    
\bibitem{edwtech} 
E. Armengaud {\it et al.} (EDELWEISS Collaboration), 
\href{https://doi.org//10.1088/1748-0221/12/08/P08010}{JINST {\bf 12}, P08010 (2017)}, arXiv:1706.01070.

\bibitem{NbSi209Migdal}

E. Armengaud {\it et al} (EDELWEISS Collaboration)  \href{https://doi.org//10.1103/PhysRevD.106.062004}{Phys. Rev. D {\bf106}, 062004 (2022)}  arXiv:2203.03993

\bibitem{nbsi-ltd}
S. Marnieros {\it et al.} (EDELWEISS Collaboration), 
Submitted to JLTP, Special Issue for the 19th International Workshop on Low Temperature Detectors, arXiv:2201.01639.

\bibitem{Neganov} 
B. Neganov and V. Trofimov, Otkryt.\ Izobret.\ {\bf 146}, 215 (1985), USSR Patent No. 1037771.

\bibitem{Luke} 
P. N. Luke, J.\ Appl.\ Phys.\ {\bf 64}, 6858 (1988).


\bibitem{knoll}
G. F. Knoll, Radiation Detection and Measurement, 4th ed.
(John Wiley and Sons, New York, 2010).


\bibitem{outer-essig}
R. Essig, J. Pradler, M. Sholapurkar and T.-T. Yu,
\href{10.1103/physrevlett.124.021801}{Phys.\ Rev.\ Lett.\  {\bf 124}, 021801 (2020)}, arXiv:1908.10881.




\bibitem{Kavanagh:2016pyr}
B. J. Kavanagh, R. Catena and C. Kouvaris, 
\href{https://doi.org//10.1088/1475-7516/2017/01/012}{JCAP {\bf 1701} (2017) 012.}

\bibitem{Emken:2017qmp}
T. Emken, C. Kouvaris. \href{https://doi.org//10.1088/1475-7516/2017/10/031}{JCAP {\bf 10} (2017) 031}

\bibitem{Mahdawi:2018euy}
M. Shafi Mahdawi, Glennys R. Farrar, arXiv:1712.01170

\bibitem{Hooper:2018bfw}
D. Hooper and S. D. McDermott,
\href{https://doi.org//10.1103/PhysRevD.97.115006}{Phys. Rev. D {\bf 97}, 115006 (2018)}
arXiv:1802.03025

\bibitem{Akerib:2018hck} 
D.~S.~Akerib {\it et al.} (LUX Collaboration),\href{https://doi.org//10.1103/physrevlett.122.131301}{Phys.\ Rev.\ Lett.\ {\bf 122}, 131301 (2019)} arXiv:1811.11241.

\bibitem{xenon-migdal}
E. Aprile {\it et al.} (XENON Collaboration),
\href{https://doi.org//10.1103/PhysRevLett.123.241803}{Phys.\ Rev.\ Lett.\ {\bf 123}, 241803}, arXiv:1907.12771.

\bibitem{cresst3} 
A. Abdelhameed {\it et al}. (CRESST),
\href{https://doi.org//10.1103/PhysRevD.100.102002}	{Phys. Rev. D {\bf 100}, 102002 (2019)}
arXiv:1904.00498.


\bibitem{milestonemodel}
J. Alexander et al., Dark Sectors 2016 Workshop: Community Report, arXiv:1608.08632


\bibitem{hemt} 
A. Juillard {\it et al}, \href{https://doi.org//10.1007/s10909-019-02269-5}{J. Low Temp. Phys. {\bf 199}, 798 (2020)}.





\end{thebibliography}

\nolinenumbers

\end{document}